\documentclass[conference]{IEEEtran}

\IEEEoverridecommandlockouts

\newcommand\size{0.8}

\usepackage{cite}
\usepackage{balance}
\usepackage{textcomp}
\usepackage{url}
\usepackage[labelfont=bf,textfont=bf]{caption}
\usepackage{subcaption}
\usepackage{graphicx}
\usepackage{xcolor}
\usepackage{xspace}
\usepackage{array}
\usepackage[noend]{algpseudocode}
\usepackage{dblfloatfix}
\usepackage{balance}
\usepackage{algorithm}
\usepackage{mathtools}
\usepackage{amsmath}

\newcommand{\shortonly}[1]{\empty{}}
\newcommand{\longonly}[1]{\empty{#1}}

\algblock[lock]{Lock}{Unlock}

\newcommand{\name}[0]{DynaHash\xspace}
\newcommand{\static}[0]{StaticHash\xspace}

\newcommand{\refalgorithm}[1]{Algorithm~\ref{#1}}

\newcommand{\reffigure}[1]{Figure~\ref{#1}}
\newcommand{\refsection}[1]{Section~\ref{#1}}

\newcommand{\refappendix}[1]{Appendix}

\newcommand{\zap}[1]{}

\newcommand{\btrees}{B$^+$-trees\xspace}

\newcommand{\revise}[1]{{\color{black}#1}\xspace}

\def\BibTeX{{\rm B\kern-.05em{\sc i\kern-.025em b}\kern-.08em
		T\kern-.1667em\lower.7ex\hbox{E}\kern-.125emX}}

\begin{document}
	
\title{\name: Efficient Data Rebalancing in Apache AsterixDB \longonly{(Extended Version)}}

\author{
	\IEEEauthorblockN{Chen Luo\textsuperscript{*}}
	\IEEEauthorblockA{
		\thanks{\textsuperscript{*} Work done while the author was at University of California, Irvine.}
		\textit{Snowflake Inc.} \\
		chen.luo@snowflake.com}
	\and
	\IEEEauthorblockN{Michael J. Carey}
	\IEEEauthorblockA{\textit{University of California, Irvine} \\
		mjcarey@ics.uci.edu}
	}

\maketitle

\begin{abstract}
	Parallel shared-nothing data management systems have been widely used
	to exploit a cluster of machines for efficient and scalable data processing.
	When a cluster needs to be dynamically scaled in or out, data must be efficiently rebalanced.
	Ideally, data rebalancing should have a low data movement cost, incur a small overhead on data ingestion and query processing,
	and be performed online without blocking reads or writes.
	However, existing parallel data management systems often exhibit certain limitations and drawbacks in terms of efficient data rebalancing.
	
	In this paper, we introduce \name, an efficient data rebalancing approach that combines \emph{dyna}mic bucketing with extendible \emph{hash}ing for shared-nothing OLAP-style parallel data management systems.
	\name dynamically partitions the records into a number of buckets using extendible
	hashing to achieve good a load balance with small rebalancing costs.
	We further describe an end-to-end implementation of the proposed approach inside an open-source
	Big Data Management System (BDMS), Apache AsterixDB.
	Our implementation exploits the out-of-place update design of LSM-trees to
	efficiently rebalance data without blocking concurrent reads and writes.
	Finally, we have conducted performance experiments using the TPC-H benchmark and we present the results here.
\end{abstract}

\section{Introduction}
The coming end of Moore's law and the information age have led
data management systems to exploit clusters of machines to process
large amounts of fast growing data.
As a result, parallel shared-nothing data management systems have become widely used due to their scalability.
In a shared-nothing parallel data management system,
records are partitioned across a cluster of nodes that communicate with each other via an interconnection network~\cite{parallel-db1992}.
The shared-nothing parallel architecture enables these systems to be horizontally scaled as the number of nodes increases~\cite{parallel-db1992}.

Early parallel data management systems~\cite{bubba1990,gamma1990} generally assumed that the cluster of nodes is relatively static.
However, this assumption is no longer true. It is desirable to dynamically adjust the cluster size for a number of reasons.
For example, it can be economical to dynamically scale the cluster in and out as the workload changes, especially in the era of cloud computing.
Moreover, as stored data accumulates over time, the cluster also needs to be scaled out to better serve the query workloads.
In order to scale a cluster in or out, the stored records must be rebalanced so that they can be repartitioned to the new set of nodes.
Ideally, rebalance operations should result in a near-perfect load balance, impose a low overhead on regular database operations, and involve a small data movement cost.
Moreover, rebalancing must be performed online so that reads and writes are not blocked.

In this paper, we focus on data rebalancing for
shared-nothing parallel data management systems aimed at analytical (OLAP) workloads.
Even though many parallel data management systems today have implemented various data rebalancing functionalities,
the existing implementations often exhibit certain limitations or drawbacks for OLAP data management.
Data management systems that support OLTP workloads~\cite{hbase,mongodb,bigtable2008,tidb2020,cockroachdb2020} often perform fine-grained range partitioning to enable efficient data rebalancing.
However, this is not suitable for \revise{shared-nothing OLAP systems} due to the potential query load imbalance caused by range skews.
Existing OLAP systems either rely on the \revise{underlying shared storage for data rebalancing~\cite{snowflake2016}},
incur a large data movement cost~\cite{vertica2012},
and/or block writes during rebalancing~\cite{redshift2015}.

\textbf{Our Contributions.}
In this paper, we present \name, an efficient data rebalancing approach for OLAP-style parallel data management systems
with local secondary indexes.
The basic idea of \name is to dynamically partition the records into a set of buckets using extendible hashing~\cite{extendible-hashing1979} and to move buckets for efficient rebalancing.
By combining extendible hashing with dynamic bucketing, \name can greatly reduce the data movement cost with a minimal impact
on data ingestion and query processing performance.

As the second contribution, we describe an efficient rebalancing implementation that avoids blocking concurrent reads and writes
by exploiting the out-of-place design offered by LSM-trees~\cite{lsm1996}.
The techniques used include bucketed LSM storage, lazy secondary index cleanup, concurrency control for online rebalancing, and an effective approach to fault tolerance and recovery.
Even though some similar individual techniques have been implemented by other systems, our contribution here is to
show how to integrate them together to yield an efficient and effective rebalancing implementation.

As the last contribution, we have implemented all of the proposed techniques inside Apache AsterixDB~\cite{asterixdb-web}
and have carried out extensive experiments on the TPC-H benchmark~\cite{tpch}  to evaluate the effectiveness of the techniques.
The experimental results show that the proposed rebalancing approach in \name
significantly reduces the rebalance cost with just a small overhead on query and ingestion performance.
\revise{It should be noted that while our approach has been implemented in an LSM-based row store,
the design itself can be naturally generalized to shared-nothing OLAP systems based on column-stores since those systems have generally adopted the same out-of-place update design for their data~\cite{hana,vertica2012,cstore}.}

The remainder of this paper is organized as follows.
\refsection{sec:background} discusses background information and related work.
\refsection{sec:architecture} presents an overview of our proposed rebalancing approach.
\refsection{sec:lsm-storage} describes how to store buckets efficiently on a single node.
\refsection{sec:asterixdb-rebalancing} presents the detailed design and implementation of the rebalance operation.
\refsection{sec:expr} experimentally evaluates the proposed techniques.
Finally, \refsection{sec:conclusion} concludes the paper.

\section{Background}
\label{sec:background}

\subsection{Data Rebalancing}
\label{sec:rebalancing}
To exploit the parallelism provided by a cluster of nodes, the records of a dataset must be distributed to each node using a partitioning function.
A partitioning function deterministically assigns each record to a node based on its partitioning key.
Example partitioning functions include range partitioning and hash partitioning\footnote{There could be other partitioning functions in practice, such as round-robin partitioning and random partitioning. However, we do not consider them here because those partitioning functions are not deterministic.}.
Range partitioning divides the key space into a set of ranges, each of which is assigned to a node.
In contrast, hash partitioning operates on the hashed keys to achieve a better load balance.

When the cluster needs to be scaled in or out, its datasets must be repartitioned through a rebalance process.
In general, rebalancing has three important trade-offs, i.e., the load balance, the rebalance cost, and the normal operation overhead.
The load balance measures how evenly the data is distributed across different nodes.
This directly impacts query performance, as in a shared-nothing system the query time is bottlenecked by the slowest node.
The rebalance cost measures how much of the data needs to be accessed and moved during rebalancing.
Finally, the normal operation overhead measures any extra overhead for normal read and write operations in order to
support the needs of the rebalance operation.

Rebalancing data changes its partitioning function.
Depending on how the partitioning function changes, existing rebalancing schemes can be classified as either \emph{global} or \emph{local}.
A global rebalancing scheme repartitions (nearly) all records of a dataset when the cluster changes.
This generally leads to a near-perfect load balance and a small normal operation overhead but at a very high rebalance cost.
For example, with range partitioning, a global rebalancing scheme can recompute the key range of each node based on the new cluster size
and then repartition all records based on the new ranges.

In contrast to global rebalancing, a local rebalancing scheme only changes the partitioning function ``locally''
so that only a small portion of the records, generally proportional to the affected nodes, are moved.
This reduces the rebalancing cost, but generally leads to a worse load balance and a higher normal operation overhead.
Commonly used local rebalancing schemes include \emph{static bucketing}, \emph{dynamic bucketing}, and \emph{consistent hashing}.
In static bucketing, the key space is pre-partitioned into a fixed number of buckets, each of which is assigned to a node through a directory.
During rebalancing, only a small number of affected buckets are moved to new nodes, which significantly reduces the rebalance cost.
Dynamic bucketing further extends the usability of static bucketing by dynamically splitting or merging buckets as the dataset size grows or shrinks.
Finally, consistent hashing eliminates the overhead of the global directory by organizing the (hashed) key space into a ring structure and letting each node serve a key range\footnote{The range partitioning counterpart of consistent hashing is rarely used in practice because of the potential for range skews. Thus, that scheme is not considered here.}.
When a node is added or removed, its key range is adjusted locally based on its next neighbor node.
In general, consistent hashing is more suitable for a large peer-to-peer architecture since it does not require a global directory.
In contrast, dynamic bucketing works naturally with a more centralized (primary-secondary)
architecture where the bucket assignment information is managed by the coordinator.

\subsection{Log-Structured Merge Trees}
\label{sec:background-lsm}
The LSM-tree~\cite{lsm1996} is a persistent index structure optimized for write-intensive workloads.
The LSM-tree adopts an out-of-place update design by always buffering writes into a memory component and appending records to a transaction log for durability.
Whenever the memory component is full, writes are flushed to disk to form an immutable disk component.
Multiple disk components are periodically merged together to form a larger one, according a pre-defined merge policy.

A query over an LSM-tree has to reconcile the entries with identical keys from multiple components, as entries from newer components override those from older components.
A range query searches all components simultaneously using a priority queue to perform reconciliation.
A point lookup query simply searches all components from newest to oldest until the first match is found.
To speed up point lookups, a common optimization is to build Bloom filters~\cite{bloom-filter1970} over the sets of keys stored in disk components.

\subsection{Apache AsterixDB}
\label{sec:asterixdb}
Apache AsterixDB~\cite{asterixdb-web,asterixdb2014,asterixdb2019} is an open-source Big Data Management System (BDMS) that aims to manage massive amounts of semi-structured (e.g., JSON) data efficiently.
AsterixDB uses a shared-nothing parallel architecture with local secondary indexes for OLAP-style workloads.
An AsterixDB cluster contains a Cluster Controller (CC) that serves as the coordinator and multiple Node Controllers (NCs)
that perform data processing tasks.
Each NC has multiple partitions to exploit the parallelism of modern hardware.
{A query in AsterixDB is compiled and optimized by the CC into a Hyracks job~\cite{hyracks2011} that is then executed by the NCs.
To support efficient data ingestion, AsterixDB provides data feeds~\cite{asterixdb-feed2015}, which are long-running jobs that constantly ingest external data into AsterixDB.}

The records of an AsterixDB dataset are hash-partitioned based on their primary keys across multiple NC partitions.
Each dataset partition is managed by an LSM-based storage engine~\cite{asterixdb-storage2014}, including a primary index, a primary key index, and multiple local secondary indexes.
The primary index stores records indexed by primary keys, and the primary key index stores primary keys only.
The primary key index is built to support COUNT(*) style queries and uniqueness checks efficiently~\cite{lsm-storage2019} since it is much smaller than the primary index.
Secondary indexes use the composition of the secondary key and the primary key as their index keys.
AsterixDB supports LSM-based \btrees, R-trees, and inverted indexes using a generic LSM-ification framework
that can convert an in-place index into an LSM-based index.
Each LSM-tree uses a tiering-like merge policy to merge its disk components.
AsterixDB uses a record-level transaction model to ensure that indexes are kept consistent within each partition.

AsterixDB currently uses a global rebalancing scheme with hash partitioning.
Given a cluster with $N$ partitions, AsterixDB assigns each record with key $K$ to the $hash(K) \textrm{ mod } N$ partition.
When the cluster size changes, the partitioning function is recomputed so that the records of a dataset
are redistributed to the new set of nodes.
This approach leads to a near-perfect load balance with a minimum normal operation overhead, but the rebalance cost is very high since nearly all records need to be moved during rebalancing.
In this work, we explore alternative data rebalancing schemes to make better trade-offs among these three costs.

\subsection{Related Work}
\label{sec:related-work}
\textbf{Rebalancing in Parallel Data Management Systems.}
Nearly all parallel data management systems today have implemented some form of rebalancing.
Here we discuss rebalancing in some representative systems based on the taxonomy of \refsection{sec:rebalancing}.

For OLTP-style systems, Google's Bigtable~\cite{bigtable2008} and its open-source cousin HBase~\cite{hbase} use dynamic bucketing with a shared-data architecture.
Since their underlying distributed storage systems already support rebalancing of immutable data blocks,
Bigtable and HBase only need to manage their in-memory data during rebalancing.
Dynamo~\cite{dynamo2007} and its open-source cousin Cassandra~\cite{cassandra2010} are shared-nothing systems that
use consistent hashing.
Cassandra further introduces the concept of virtual nodes to achieve a better load balance;
the basic idea is to let each node use multiple virtual nodes to manage multiple key ranges.
Couchbase~\cite{couchbase2016} and Oracle NoSQL Database~\cite{oracle-nosql} are shared-nothing systems
that use static bucketing with hash partitioning.
Both systems set the number of buckets to a relatively high number.
Couchbase sets this number to 1024 by default,
while Oracle NoSQL Database recommends that each node (in the expected largest cluster) should have 10 to 20 buckets.
Minhas et al.~\cite{scale-out2012} applied a similar static bucketing approach to enable efficient scaling for VoltDB~\cite{voltdb}.
MongoDB~\cite{mongodb},  TiDB~\cite{tidb2020}, WattDB~\cite{physilogical2015}, and CockroachDB~\cite{cockroachdb2020}
each use range-partitioned dynamic bucketing with a very small bucket size, e.g., 64MB.
\revise{Such range partitioning is suitable for OLTP workloads since each transaction only accesses a small number of (usually one)  partitions.
However, it may not be suitable for shared-nothing OLAP systems due to the potential range skews and the fact that queries will often access most partitions.
OLTP systems typically use global secondary indexes due to the high selectiveness of OLTP queries.}

For OLAP-style systems, \revise{Snowflake~\cite{snowflake2016} decouples storage from compute and completely relies on
the underlying shared storage for data rebalancing.
It is worth noting that Snowflake also uses consistent hashing to cache immutable data files into compute nodes.
However, this is much simpler than data rebalancing considered here because the data files are immutable and one can simply invalidate the cache when the cluster changes.}
Vertica~\cite{vertica2012} is a shared-nothing system
that uses global rebalancing with hash partitioning to achieve a better load balance.
Redshift~\cite{redshift2015} is shared-nothing and supports both global rebalancing and static bucketing with hash partitioning.
However, Redshift does not support concurrent writes during rebalancing.
{Moreover, it directly uses buckets (called ``node slices'' in Redshift) as its parallelism unit, which can undesirably change the node parallelism
	after rebalancing.
}

\textbf{Elastic OLTP Databases.}
Due to the importance and wide adoption of parallel OLTP database systems, significant effort has been devoted to making them elastic.
Live migration techniques~\cite{albatross2011,zephyr,squall2015} enable OLTP databases to be migrated without blocking ongoing transactions.
E-Store~\cite{estore2014} uses fine-grained partitioning to elastically scale parallel databases.
\revise{However, a key difference is that these research efforts mainly focus on ACID transactions with fine-grained range partitioning, global secondary indexes, and small queries, while our work focuses on how to rebalance datasets efficiently for OLAP-style (i.e., analytical query-oriented) systems.}

\textbf{Distributed Access Methods.}
To efficiently query data stored in a cluster of nodes, a number of distributed access methods have been proposed as well.
The basic idea is to distribute an access method efficiently over a cluster of nodes, potentially in a peer-to-peer setting, to 
support efficient read and write operations~\cite{distributed-extendible-hashing1983,lh1993,distribute-search-tree1994, distributed-btree2008},
The difference between these access methods and our work is that we focus on rebalancing for OLAP systems
rather than on a single access method with simple key-value interfaces.

\textbf{LSM-trees.}
For data storage in modern systems,
a large number of improvements have been proposed to optimize the LSM-tree~\cite{lsm1996},
including~\cite{lsm-tuple2020,monkey2017,dostoevsky2018,lsm-bush2019,lsm-memory-short2020, lsm-storage2019,lsm-stability2019,lsm-memory2020,umzi2019,rosetta2020,lsm-secondary2018,lethe2020}.
We refer readers to a recent survey~\cite{lsm-survey2020} for a more detailed description of these LSM-tree improvements.
These LSM-tree improvements have focused on single node settings.
In contrast, in this work, we focus on their role in a parallel shared-nothing architecture and exploit the LSM-tree's out-of-place design to support efficient data rebalancing with concurrent reads and writes.

\section{Approach Overview}
\label{sec:architecture}
As mentioned in \refsection{sec:rebalancing}, rebalancing involves three important trade-offs, i.e., the load balance, the rebalance cost, and the normal operation overhead.
Our goal is to achieve a good load balance with a small rebalance cost and low normal operation overhead.
In this section, we provide a high-level overview of \name based on the following design choices.

\textbf{Range Partitioning vs. Hash Partitioning.}
\revise{Most shared-nothing OLAP systems prefer hash partitioning to achieve a good load balance because the data will be nearly uniformly distributed and queries generally access most partitions.
In contrast, range partitioning is often unsuitable for shared-nothing OLAP systems due to the potential for range skews.
Thus, we choose hash partitioning to use here as well.}

\textbf{Global Rebalancing vs. Local Rebalancing.}
Although global rebalancing achieves a near-perfect load balance, it incurs a very large rebalance cost since most records have to be moved during rebalancing.
Since our goal is to reduce the rebalance cost, we prefer to use a local rebalancing scheme.
Among the three local rebalancing schemes (\refsection{sec:rebalancing}),
dynamic bucketing dominates static bucketing by elastically adjusting the number of buckets as data accumulates.
Dynamic bucketing is also preferable to consistent hashing since most parallel OLAP systems (ours included)
adopt a primary-secondary architecture.
Based on these considerations, a natural choice is to use dynamic bucketing here.

\begin{figure}
	\centering
	\includegraphics[width=\linewidth]{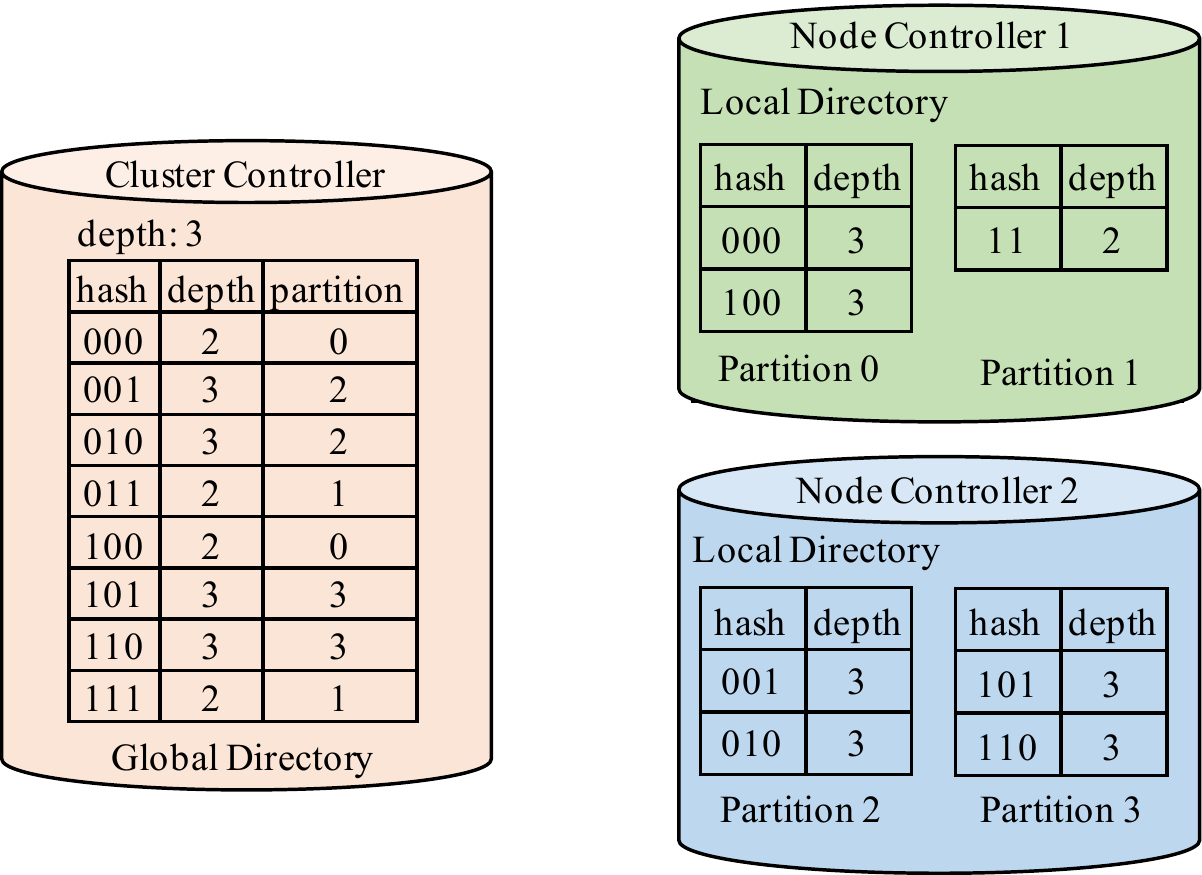}
	\caption{Example Architecture for \name}
	\label{fig:architecture}
\end{figure}

\textbf{Combining Hash Partitioning with Dynamic Bucketing.}
The last choice we face is how to combine hash partitioning with dynamic bucketing.
One natural design would be to range-partition the hashed key space into multiple buckets.
Though this solution works, hashing actually provides opportunities for a more efficient design.
Since hashed keys are \revise{generally} uniformly distributed, one can use an extendible
hashing approach~\cite{extendible-hashing1979} to partition the key space into multiple buckets.
\reffigure{fig:architecture} illustrates the resulting architecture based on this idea with one Cluster Controller (CC) and two Node Controllers (NCs).
Each NC further has two storage partitions.
In order to distribute the records of a dataset to these four partitions, the hash key space is divided into multiple buckets.
A bucket is defined by taking the $d$ low-order bits of the hash function, where $d$ is called the depth of this bucket.
When a bucket becomes too large, it is split into two smaller buckets by taking one more hash bit,
which thus increments the depth~\cite{extendible-hashing1979}.
A rebalance operation can now move only some affected buckets to new partitions,
greatly reducing the rebalance cost.

As shown in \reffigure{fig:architecture}, we use a \emph{global directory} stored at the CC to map buckets to partitions.
This directory has a global depth $D$, which is the maximum number of bits used in all buckets.
Thus, the size of this directory is $2^D$.
Note that in \reffigure{fig:architecture} the two hash values $011$ and $111$ currently correspond to the same bucket $11$.
To locate where a given key $K$ is stored, one simply needs to look in
the global directory using the $D$ low-order bits of $K$'s hash value,
where $D$ is the depth of the global directory.
During query compilation, each query creates an immutable copy of the global directory that it uses throughout query processing.
Similarly, a data feed, i.e., a data ingestion job, also employs an immutable copy of the global directory in order to distribute the incoming records of a dataset to the correct NC partitions.

We further use a \emph{local directory} at each partition to keep track of the assigned buckets.
To simplify bucket splits, the global directory can be updated lazily before rebalancing is performed.
For example, in \reffigure{fig:architecture}, the bucket $00$ has already been split into two buckets $000$ and $100$ at partition 0,
but the global directory has not been updated yet.
This does not impact the correctness of the global directory since it can still correctly route all keys to the right partitions.

Though the basic design in our rebalancing approach of relatively straightforward, two key challenges must be addressed.
First, how can we store (i.e., physically organize) the multiple buckets within each partition to enable efficient rebalancing with low normal operation overhead?
Second, how can we efficiently rebalance buckets while supporting both concurrent reads and writes?
In the next two sections we detail our solutions to these two key challenges.

\section{LSM Storage for Buckets}
\label{sec:lsm-storage}
In this section, we discuss how to efficiently store multiple buckets in each partition.
For efficient rebalancing, when a bucket needs to be moved out of a partition, it is desirable to only access the records in this bucket.
If range partitioning were used and no secondary indexes were built, storing records in their primary key order would naturally satisfy this property
since records in each bucket would be grouped together.
However, with hash partitioning, the primary key order is no longer the same as the bucket order
since records are bucketed based on hashed keys.
Moreover, secondary indexes also complicate this problem because their entries are ordered by secondary keys, not by primary keys.

\textbf{Storage Options.}
In general, since hash partitioning is used, there are three storage options for buckets in each partition:
\begin{itemize}
	\item Option 1: Store entries in their original key order together in one LSM-tree index.
	\item Option 2: Store entries in their bucket order in one LSM-tree index. Within each bucket, store its entries in their original key order.
	\item Option 3: Store the entries in each bucket in a separate LSM-tree index structure.
	Within each LSM-tree, store the entries in their original key order.
\end{itemize}

\begin{figure}
	\centering
	\includegraphics[width=\linewidth]{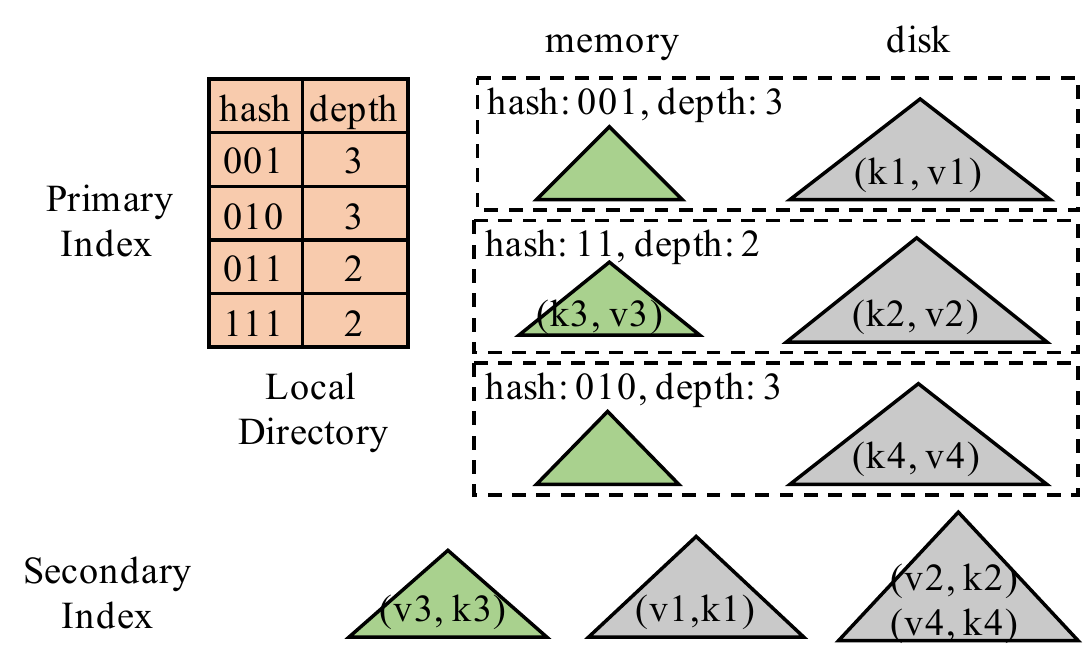}
	\caption{{An Example Partition with a Bucketed Primary Index and a Secondary Index}}
	\label{fig:bucketed-lsm}
\end{figure}

Let us first consider the trade-offs for the primary index.
Option 1 incurs no overhead on reads and writes, but it incurs a large overhead on rebalancing since moving a bucket must scan all entries, including those from other buckets.
Options 2 and 3 both reduce the rebalancing overhead since the records within each bucket are physically grouped together.
Moreover, Option 3 provides more flexibility for splitting buckets and deleting buckets after rebalancing.
However, Options 2 and 3 both induce some overhead on short primary index scans since each query must search all buckets.
Since short primary key-order scans are not common in OLAP-style systems, we choose to optimize the rebalancing performance by choosing Option 3 for the primary index of a dataset.
For secondary indexes, an important difference is that they do not have to be read during rebalancing as they can be rebuilt on-the-fly at their destination.
In order not to incur too much normal runtime overhead on secondary index queries, we choose to use Option 1 for secondary indexes.
{\reffigure{fig:bucketed-lsm} shows an example of a dataset partition with a primary index and a secondary index.
Here we denote each record as a key-value pair, and the secondary index is built on the value field.
The primary index uses the bucketed LSM-tree design, which is further described below, to store buckets separately.
The secondary index uses a traditional LSM-tree design to store all buckets together.}

\textbf{Bucketed LSM-tree Design.}
Based on these basic design decisions,
we introduce a bucketed LSM-tree design for efficiently storing multiple buckets in the primary index.
As shown in \reffigure{fig:bucketed-lsm},
each bucket can be viewed as a separate LSM-tree with a memory component and multiple disk components.
{We use reference counting for concurrency handling.
That is, whenever a bucket, a memory component, or a disk component is accessed, the reader or writer increments a reference count
so that the accessed entity cannot be destroyed (reclaimed) until the access completes.}
All flushes and merges are performed within each bucket.
All buckets are coordinated using a local directory, as mentioned in \refsection{sec:architecture}.
Note that in \reffigure{fig:bucketed-lsm}, hashes $011$ and $111$ correspond to the same bucket $11$ with depth 2.

\textbf{Data Ingestion and Query Processing.}
A bucketed LSM-tree provides the same set of interfaces as a traditional LSM-tree.
A write operation, including inserts, deletes, and updates, first checks the local directory using the hash value of the key
to locate which bucket the entry belongs to and then adds the entry to that bucket.
Similarly, a point lookup query only searches its target bucket, located via the local directory, to get the entry.
A primary key range scan query, however, must search all buckets.
There are two approaches to process such a range scan query.
The first approach is to scan each bucket separately.
This will incur no additional overhead compared to the traditional LSM-tree design,
but the returned entries will no longer be sorted on the primary key.
The second approach is to use a priority queue to \revise{merge} the entries returned from all buckets together.
This approach provides the same interface as the traditional LSM-tree design by returning sorted results,
but will incur a larger search overhead due to the additional merge-sort step.
To decide which approach should be used, we have introduced an optimization rule in AsterixDB as follows.
By default, the first approach is used to avoid the merge-sort overhead.
However, if primary key order is required by subsequent query operators,
e.g., a user-specified order-by clause or a groupby operator on a prefix of the primary key,
the second approach will be used to avoid the subsequent sort overhead.
{Finally, it should be noted that the bucketed LSM-tree design does not change the processing of a secondary index query,
which simply searches the secondary index to fetch a list of primary keys and then uses them to fetch records from the (bucketed)
primary index.
}

\begin{figure}
	\centering
	\includegraphics[width=1\linewidth]{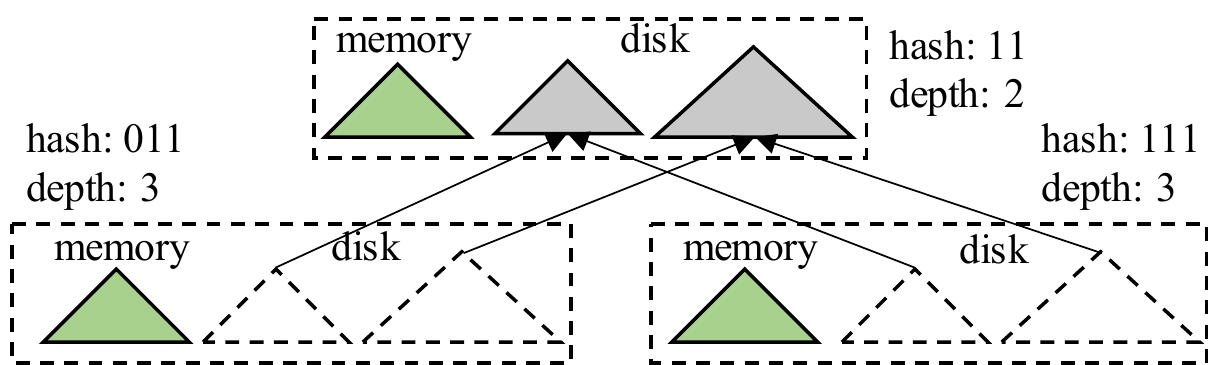}
	\caption{Bucket Split Example}
	\label{fig:split}
\end{figure}

\textbf{Efficient Bucket Splits.}
When a bucket becomes too large, it is split into two smaller buckets by using one more hash bit.
A straightforward implementation of this would be to build two smaller LSM-trees based the original bucket.
However, this approach would not only cause additional write amplification, but might also block reads and writes for a long time.
Here we describe a more efficient bucket splitting approach to address these issues.

The pseudocode for splitting a bucket $B$ is depicted in \refalgorithm{alg:bucket-split}.
The \textsc{Split} function first stops initiating new component merges for $B$ and waits for all existing merges to finish.
$B$'s memory component is then asynchronously flushed to disk without blocking writes (line 5).
{After the flush completes, the bucket $B$ is locked to temporarily block
new readers and writers so that $B$ can be safely split (lines 6 to 10).}
Since some writes may have entered its memory component after the last asynchronous flush,
$B$'s memory component is now flushed synchronously to persist these writes.
(It should be noted that AsterixDB uses a no-steal buffer management policy,
meaning that a memory component is only flushed after all active writers have completed.)
Two new buckets $B_1$ and $B_2$, whose disk components refer to the disk components of $B$, are then created.
An example is shown in \reffigure{fig:split}. For each disk component of the splitting bucket $11$,
we create two new \emph{reference disk components} in buckets $011$ and $111$ respectively.
A reference disk component does not store any data; instead, it only points to a real disk component.
All queries accessing data through a reference disk component must perform an additional filtering step based on the bucket's hash value to make sure that only the entries truly belonging to this bucket are accessed.
Thus, the actual creation of the new disk components of $B_1$ and $B_2$ are effectively postponed until the next round of merges.
Finally, a directory metadata file that stores valid buckets is forced to disk, indicating that the split operation is now complete (line 9),
and the old bucket $B$ is reclaimed automatically when its reference count becomes 0.
Upon recovery, the directory metadata file is used to determine the valid buckets. All invalid (partially split) buckets will be cleaned up automatically.

\algblockdefx[lock]{Lock}{Unlock}
{Lock $B$}{Unlock $B$}

\begin{algorithm}
	\small
	\centering
	\begin{algorithmic}[1]
		\State	$B$ $\gets$ the bucket to be split
		\Function{Split}{B}
			\State{Pause scheduling merges for $B$}
			\State{Wait for $B$'s merges to finish}
			\State{Asynchronously flush $B$'s memory component}
			\Lock
			\State{Synchronously flush $B$'s memory component}
			\State{Create two buckets $B_1$ and $B_2$ that refer to $B$}
			\State{Force a directory metadata file to disk}
			\Unlock
			\State{Resume scheduling merges for $B$}
		\EndFunction
	\end{algorithmic}
	\caption{Pseudo Code for Bucket Split}
	\label{alg:bucket-split}
\end{algorithm}

To simplify synchronization with the CC, bucket splits are performed at each partition locally without notifying the CC.
Instead, the global directory at the CC is only refreshed when a rebalance operation starts, as we will discuss below.
This design greatly simplifies the role of the CC since it does not have to know about the existence of bucket splits.

\section{Efficient Data Rebalancing}
\label{sec:asterixdb-rebalancing}
Having considered how to store multiple buckets efficiently, we now discuss how to efficiently rebalance data while supporting concurrent reads and writes.
In AsterixDB, data rebalancing is triggered via a system API after some nodes have been added or before some nodes are removed.
In general, a rebalance operation contains three phases, namely initialization, data movement, and finalization.
During the initialization phase, all nodes perform some preparation tasks for subsequent data movement.
The data movement phase transfers some of the records of a dataset, including concurrent writes, to their new partitions.
Finally, during the finalization phase, all nodes unanimously commit or abort the rebalance operation depending on its outcome,
and some cleanup work is performed.
It should be noted that a rebalance operation may fail for various reasons.
If a rebalance operation fails, the produced intermediate results must be cleaned up correctly.
In the remainder of this section, we discuss the three phases in detail as well as how to handle rebalance failures.

\subsection{Initialization Phase}
When a rebalance operation starts, the CC forces a BEGIN log record indicating that a rebalance operation has started.
The CC further decides which buckets should be moved to which partitions by computing a new global directory based on the new set of nodes.
In addition, all NCs must perform some preparation tasks in order to support concurrent updates.
The key challenge here is that AsterixDB only supports a very simple record-level transaction model.
If full ACID transactions were supported by AsterixDB, then the rebalance operation could potentially be implemented using a transaction,
which would automatically provide concurrency control for reads and writes.
Instead, we must design a customized concurrency control protocol.

\textbf{Computing the Global Directory.}
Recall from \refsection{sec:lsm-storage} that buckets splits are performed at each node locally, without notifying the CC.
Thus, in order to compute the new global directory, the CC contacts all NCs to get their latest local directories.
Moreover, bucket splits for this dataset at each NC are disabled until the rebalance completes.
Note that since buckets may have different sizes, it is straightforward to show that an optimal algorithm to maximize the load balance is NP-hard by considering the partition problem\footnote{The goal of the partition problem is to partition a multiset $S$ of positive integers into two subsets $S_1$ and $S_2$ such that the difference between the sum of elements in $S_1$ and the sum of elements in $S_2$ is minimized~\cite{partition}.}.

\begin{algorithm}
	\small
	\centering
	\begin{algorithmic}[1]
		\Function{Balance}{}
		\For{each unassigned bucket $B$}
		\State{Assign $B$ to the least loaded partition}
		\EndFor
		\While{$true$}
		\State{$P_{max} \gets$ the most loaded partition}
		\State{$B\gets$ the smallest bucket in $P_{max}$}
		\State{$P_{min} \gets$ the least loaded partition}
		\If{{$abs( (|P_{max}| - |B|) - (|P_{min}| + |B|) ) < |P_{max}|-|P_{min}|$}}
		\State{Assign $B$ to $P_{min}$}
		\Else
		\State{\textbf{break}}
		\EndIf
		\EndWhile
		\EndFunction
	\end{algorithmic}
	\caption{Pseudo Code for Computing Global Directory}
	\label{alg:bucket-balance}
\end{algorithm}

To compute the new global directory efficiently, we use the greedy algorithm shown in \refalgorithm{alg:bucket-balance}.
To describe this algorithm, we first introduce some useful concepts.
Given a directory with global depth $D$ and a bucket $B$ with depth $d$, we define the \emph{normalized size} of the bucket $B$ 
(denoted as $|B|$) as $2^{D-d}$.
Given a partition $P$ or a node $N$, we denote $|P|$ or $|N|$ as the sum of the normalized size of $P$'s buckets and $N$'s buckets, respectively.
Given two partitions, $P_1$ on node $N_1$ and $P_2$ on node $N_2$, $P_1$ is said to be more loaded than if $|P_1|$ is larger than $|P_2|$,
or $|N_1|$ is larger than $|N_2|$ if $|P_1|$ equals $|P_2|$.
The \textsc{Balance} function first assigns the unassigned buckets (buckets being displaced due to node removals)
to the least loaded partitions (lines 2-4).
After all such buckets are assigned, the algorithm balances the bucket assignment using a series of iterations (lines 4-11).
In each iteration, it tries to assign the smallest bucket $B$ from the most loaded partition $P_{max}$ to the least loaded partition $P_{min}$ (lines 8-11).
If this assignment reduces the difference between the normalized sizes of $P_{max}$ and $P_{min}$, the assignment is then performed;
otherwise, the algorithm terminates.
It is possible to incorporate other heuristics to define the load order among partitions.
For example, one could further consider the total storage size of a partition, including all datasets.
We leave the exploration of this direction as future work.

\textbf{Preparing for Concurrent Writes.}
During the rebalance operation, which may take a relatively long time to finish, some records may be updated by concurrent writers.
For each bucket that needs to be moved, these concurrent writes must still be applied to its old partition since the rebalance operation may fail.
Moreover, the concurrent writes must also be applied to the new partition to ensure that there are no lost writes if the rebalance operation succeeds.

To ensure correctness with concurrent writes, we use a concurrency control protocol that splits all the writes to a bucket based on the rebalance start time.
For all writes that happened before the rebalance operation starts, an immutable bucket snapshot is created so that it can be safely scanned.
For all writes that happen after the rebalance operation starts, their log records are replicated to the new partition so that the new partition
will not miss any writes.
It should be noted that AsterixDB only supports a very simple record-level transaction model without supporting snapshot scans.
To implement the required snapshot scan, we exploit the immutability of LSM disk components.
Specifically, the memory component of the moving bucket is flushed synchronously during the initialization phase.
This flush time is treated as the rebalance start time, and the resulting disk components
become the immutable copy of all writes that happened before the rebalance operation starts.
To reduce the blocking of concurrent writes due to the synchronous flush, the two-flush approach described in \refalgorithm{alg:bucket-split} (lines 5-7) can be used.
Specifically, one can first flush the memory component asynchronously and then use a synchronous flush to persist any leftover writes.
In this case, the rebalance start time becomes the time of the second (synchronous) flush.

\zap{
	\begin{figure}
	\centering
	\includegraphics[width=0.95\linewidth]{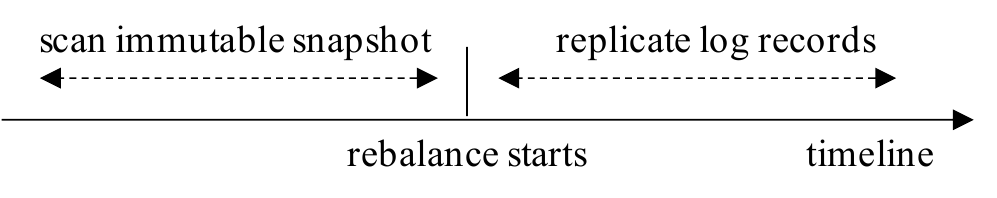}
	\caption{Concurrency Control for Writes}
	\label{fig:concurrent-update}
\end{figure}
}

\subsection{Data Movement Phase}
After the initialization phase, the rebalance operation starts to move the affected buckets to their new partitions.
This involves adding scanned records and replicated log records to both the primary index and secondary indexes at their destination partitions.
Moreover, queries must not be affected by the rebalance operation.

\textbf{Data Movement.}
By comparing the current global directory and the new global directory, it is straightforward to determine the new partition of each affected bucket.
For ease of discussion, here we first describe how to move one bucket $B$ from its old partition $P_{old}$ to its new partition $P_{new}$,
which is then extended to moving multiple buckets together.

\begin{figure}
	\centering
	\includegraphics[width=1\linewidth]{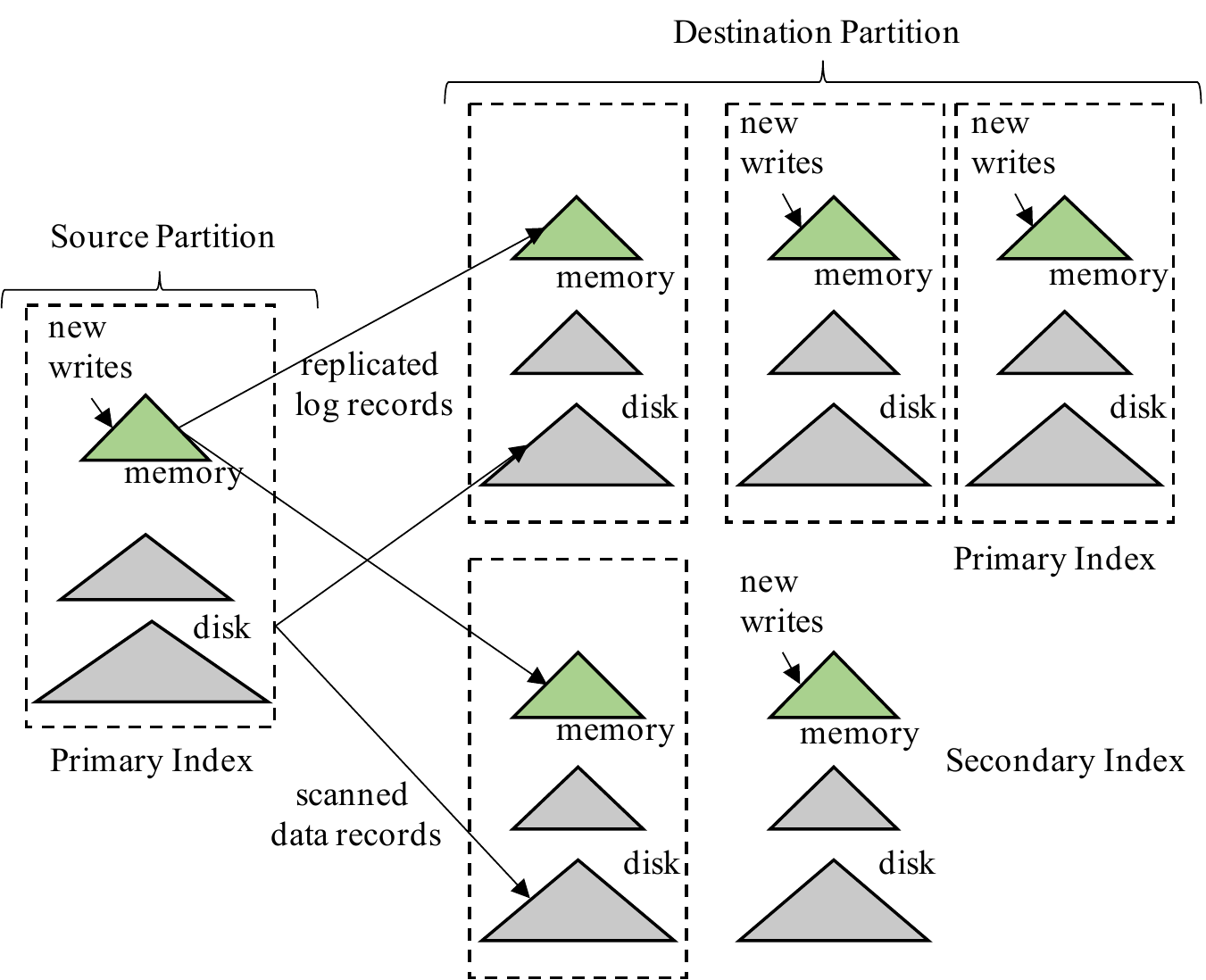}
	\caption{{Data Movement Process (One Bucket): \textmd{\emph{Scanned data records are loaded into disk components and replicated log records are inserted into memory components. These rebalance writes are stored separately from new user writes.}}}}
	\label{fig:data-movement}
\end{figure}

\reffigure{fig:data-movement} shows the basic data movement process for a single bucket with a primary index and one secondary index.
At the old partition, the primary index disk components of this bucket are scanned and the log records of any incoming writes will be replicated.
The scanned log records are used to load disk components at the new partition,
with the data represented by the replicated log records being inserted into the memory components.
In order to simplify concurrency control and recovery, the moved records are always stored separately from
local user writes at the new partition.
For a primary index that uses the bucketed LSM-tree design, the received records are simply stored in a new bucket.
For a secondary index that stores all buckets together, the received records are stored into a new list of components that
are kept invisible to queries.
This design greatly simplifies concurrency control and recovery.
The new components that store moved records will be kept invisible to queries until the rebalance completes.
Moreover, in case the rebalance operation fails, these new components can then be simply deleted to cleanup the intermediate results.
Finally, to ensure correctness, the scanned data records must be treated as being strictly older than the replicated log records.
This is achieved by placing the loaded disk component after the disk components storing replicated log records in the LSM disk component list.

It is straightforward to extend the basic data movement process to move multiple buckets at the same time.
One can simply scan multiple buckets at the same time and repartition them using the new global directory so that the scanned records can be sent to their new partitions.
As an optimization, when adding multiple buckets to a secondary index partition,
the records can be added to a single list of components instead of creating one list per bucket.
This will help to reduce the number of disk components present after the rebalance operation completes.

\textbf{Handling Concurrent Queries.}
{As mentioned before,
the data movement process shown in \reffigure{fig:data-movement} greatly simplifies the required concurrency control for queries.}
Since the moved records are stored separately from incoming user writes,
the destination's partially loaded buckets are invisible to queries until rebalancing completes.
If a query starts before a rebalance operation completes, the query accesses all buckets using a copy of the old global directory.
Otherwise, the query uses the new global directory, updated by the rebalance operation, to access all buckets.
Moreover, since accessed buckets and LSM components are reference-counted,
they can be accessed safely by the query even if a rebalance operation completes in the middle of the query.

\subsection{Finalization Phase}
\label{sec:finalize-phase}
After all the data records of moving buckets have been transferred to their new partitions, the system is ready to commit
or abort the rebalance operation depending on its outcome.
It should be noted that there could still be active log replication activities, due to concurrent writes, at this stage.
Thus, to ensure that all nodes always reach a unanimous decision,
we use a two-phase commit protocol as follows.

\textbf{Prepare Phase.}
After all data records have been moved to their new partitions, the CC initializes the prepare phase, which will temporarily
block incoming queries and writes on the rebalancing dataset.
The CC further waits for all NCs to complete their log replication and to flush the memory components that store rebalancing writes
to disk.
If all NCs succeed in doing so, i.e., they all vote yes,
the CC enters the \emph{commit} phase as discussed below.
Otherwise, the rebalance operation must be aborted and the rebalancing dataset will be left unchanged.
All incoming reads and writes will be blocked during this finalization phase.
However, this blocking is expected to be very short since the CC only waits for existing writers to complete and 
since the number of log records pending for replication is bounded.

\textbf{Commit Phase.}
Once the CC enters the commit phase, it forces a COMMIT log record to disk indicating that the rebalance operation is committed.
The CC then updates the global directory of the rebalancing dataset and notifies all NCs to install their received buckets
and cleanup the moved buckets.
To install a received bucket at a partition, 
the NC simply needs to add (i.e., register) the loaded disk components to the component lists of the primary index and secondary indexes.
To cleanup a moved bucket from the primary index of a partition, the bucket can be simply removed from the bucketed LSM-tree's local directory so that it cannot be accessed by new queries.
It should be noted that due to reference counting, the actual components of this bucket will not be deleted until the last reader exits.
To cleanup a secondary index, we use a lazy delete approach that adds the hash value and the depth of this bucket to the metadata of each LSM component.
A query then performs an additional validation check to ignore all invalid entries that belong to this moved bucket.
Thus, the cleanup of secondary index components is effectively postponed to the next round of merges.
All these operations, e.g., adding and removing buckets, are made persistent by forcing metadata files to disk.
After all NCs have completed these tasks, the CC can resume query processing and data ingestion on the rebalanced dataset.
Finally, the CC produces a DONE log record to indicate that no additional work is needed for this rebalance operation.

Based on the two-phase commit protocol,
the final outcome of the rebalance operation is determined by whether the COMMIT log record has been forced to disk successfully by the CC.
In other words, the rebalance operation is committed if the COMMIT log record has been successfully forced to disk.
Otherwise, the CC simply aborts the rebalance operation and leaves the original dataset as is. \shortonly{However, the intermediate results produced by the rebalance operation must be cleaned up carefully to ensure that the dataset remains in a consistent state. Due to space limitations, we leave the detailed discussion of failure handling to the extended version of this paper~\cite{rebalance-extended}.}

\longonly{
\begin{figure}
	\centering
	\includegraphics[width=1\linewidth]{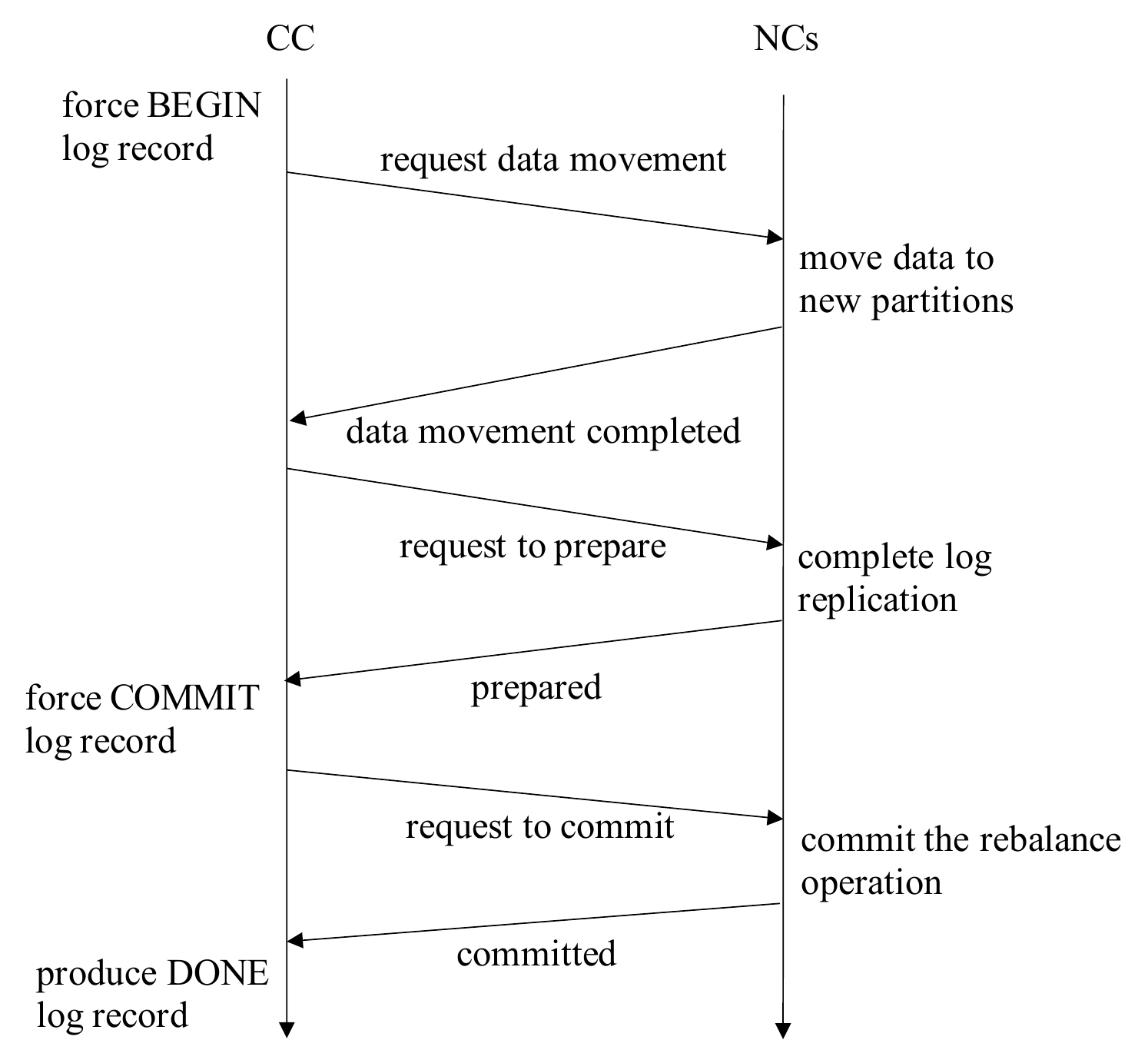}
	\caption{Rebalance Operation Timeline}
	\label{fig:timeline}
\end{figure}

\subsection{Handling Rebalance Failures}
\label{sec:rebalance-failure}
During the rebalance operation, some node(s) may potentially fail at any time.
Since node failures are expected to be rare, we simply abort the rebalance operation
if some node(s) fails before the rebalance operation commits.
However, the intermediate results produced by the rebalance operation must be cleaned up
carefully to ensure that the dataset remains in a consistent state.
Here we assume that the failed node(s) eventually recovers, i.e., no permanent node failures.
We plan to extend \name to incorporate replication~\cite{asterixdb-replication2016}
to handle permanent node failures as future work.

Before discussing how to handle various rebalance failures, we first summarize the basic timeline of a rebalance operation,
which is shown in \reffigure{fig:timeline}.
The CC first forces a BEGIN log record indicating that the rebalance operation has started.
It then requests all NCs to move their affected buckets to new partitions.
After all data movement is done, the CC enters the prepare phase by waiting for all NCs to complete log replication.
After all NCs have successfully prepared, the CC enters the commit phase by forcing a COMMIT log record
and notifying all NCs to commit this rebalance operation.
Finally, after all NCs have committed, the CC produces a DONE log record indicating that the rebalance operation can be safely forgotten.
Based on this timeline, we present a case analysis to discuss how to handle various possible rebalance failures.

\textbf{Case 1: NC fails before voting ``prepared''.}
In this case, the CC simply aborts the rebalance operation and asks all NCs {(including the failed NC after its recovery)}
to cleanup their received buckets.
Recall from \reffigure{fig:data-movement} that the received records are always added to a separate list of components.
Thus, to cleanup the received buckets, a partition can simply delete those lists of components for both the primary index and secondary indexes of the dataset.
It should be noted that cleaning up a received bucket is idempotent since cleaning up a non-existent bucket can be simply treated as a no-op.
It is thus safe to cleanup a received bucket from a partition multiple times.
After all NCs complete the cleanup task, the CC writes a DONE log record so that this rebalance operation can be safely forgotten.

\textbf{Case 2: NC fails after voting ``prepared''.}
{After the failed NC recovers, it contacts the CC to report its presence.
The NC will further receive instructions about how to handle the pending rebalance operation.}
If the rebalance operation is aborted, the NC simply cleans up the intermediate results as in Case 1.
Otherwise, the NC performs the commit tasks as in Case 4.

\textbf{Case 3: CC fails before forcing the COMMIT log record.}
After the CC recovers and sees the BEGIN log record for the rebalance operation,
it aborts the rebalance operation as in Case 1.

\textbf{Case 4: NC fails before responding ``committed''.}
The rebalance operation is committed but the CC does not know whether the NC has committed the rebalance operation or not.
When the NC recovers, the CC requests this NC to commit the rebalance operation by
adding the received buckets and cleaning up the moved buckets.
Similar to case 3, both adding the received buckets and cleaning up the moved buckets are idempotent operations,
which means it is safe to apply these operations multiple times.

\textbf{Case 5: CC fails after forcing the COMMIT log record but not the DONE log record.}
In this case, the rebalance operation is effectively committed but it is possible that some NCs have not completed the commit tasks yet.
Thus, after the CC recovers, it notifies all NCs to add received buckets and cleanup moved buckets as in Case 4.
Finally, the CC writes a DONE log record as well.

\textbf{Case 6: CC fails after the DONE log record is persisted.}
No additional task needs to be performed in this case since the DONE log record indicates that this rebalance operation has completed.

The two-phase commit protocol used in our rebalance operation has some subtle differences from the traditional two-phase commit protocol used in distributed transactions. For example, the CC forces a BEGIN log record when a rebalance operation starts and NCs always contact the CC during recovery.
This is because in AsterixDB the rebalance operation is implemented as a metadata transaction,
and only the CC can produce metadata log records.
In contrast, in traditional distributed transactions each participant can produce log records.
Because of this difference, without forcing the BEGIN log record, the CC may not know the existence of a rebalance operation if the entire cluster shuts down before the rebalance operation completes.
Similarly, the NC must always contact the CC upon recovery since the NC cannot certainly
know the status of ongoing rebalance operations.
(Contacting the CC upon recovery does not involve additional overhead since the NC must register itself with the CC for cluster management anyway.)
}

\section{Experimental Evaluation}
\label{sec:expr}
In this section, we experimentally evaluate the proposed rebalancing techniques in the context of Apache AsterixDB~\cite{asterixdb-web}.
Throughout the evaluation, we mainly focus on the three aspects of different rebalancing approaches,
namely their rebalancing performance, ingestion performance, and query performance.
We first describe the general experimental setup followed by the detailed evaluation results.

\subsection{Experimental Setup}
\label{topic3:sec:expr-setup}
\textbf{Hardware.}
All experiments were performed on a cluster of nodes with a single CC and multiple NCs on AWS.
The number of NCs ranged from 2 to 16.
The CC ran on a m5.xlarge node with 4 vCPUs, 16GB of memory, and a 500GB elastic block store (EBS).
Each NC ran on an i3.xlarge node with 4 vCPUs, 30.5GB of memory, a 950GB SSD, and a 500GB EBS.
We configured 4 partitions on each NC to exploit the parallelism provided by multiple cores.
The native SSD was used for LSM storage and the EBS was used for transactional logging.
Each NC used a thread pool with 4 threads to execute LSM flush and merge operations.
Each LSM-tree used a size-tiered merge policy with a size ratio of 1.2 throughout the experiments, which is similar to the setting used in other systems.
This policy merges a sequence of components when the total size of the younger components is 1.2 times larger than that of the oldest component in the sequence.
Within this, we allocated 26GB per node of memory for the AsterixDB instance.
The buffer cache size was set at 8GB and the memory component budget was set at 2GB.
Each memory-intensive query operator~\cite{asterixdb-mem2020},
such as sort, hash join, and hash group by, received a 128MB memory budget.
Both the disk page size and memory page size were set at 16KB.

\textbf{Workload.}
To understand the performance impact of different rebalancing approaches on OLAP-style workloads, we used the TPC-H~\cite{tpch} benchmark in our evaluation.
We built two secondary indexes, on LineItem and Orders, to enable index-only plans for certain queries.
The LineItem index contains l\_shipdate, l\_partkey, l\_suppkey, l\_extendedprice, l\_discount, and l\_quantity.
The Orders index contains o\_orderdate, o\_custkey, o\_shippriority, and o\_orderpriority.
The scale factor of the TPC-H benchmark was set to 100 times the number of NCs so that the total amount of data scales linearly
as the cluster size increases.
Thus, each NC stored 100GB of TPC-H raw data.
The primary index was compressed for better storage efficiency.
The total storage size at each NC, including compressed primary indexes and uncompressed secondary indexes, was about 130GB.

\textbf{Evaluated Rebalancing Approaches.}
We evaluated three rebalancing approaches.
The first approach that we evaluated is
AsterixDB's current global rebalancing approach with hash partitioning (called ``Hashing'') as the baseline.
This approach simply creates a new dataset that is hash-partitioned based on the new (target) set of nodes during rebalancing.
Although Hashing achieves a near perfect load balance, it incurs a very high rebalance cost and nearly doubles the dataset's
disk usage during rebalancing.
\revise{We further evaluated two variations of the proposed rebalancing approach.
The first variation, called \static, always splits a dataset into 256 buckets.
The actual number of buckets per partition ranged from 32 to 4 as the number of nodes varied from 2 to 16.
Thus, evaluating \static also shows the performance impact of the number of buckets per partition.
The second variation, called \name, dynamically splits a dataset by setting the maximum bucket size at 10GB.
After loading the TPC-H data, each partition always had 4 buckets.}

\subsection{Ingestion Performance}

\begin{figure}
	\centering
	\includegraphics[width=0.65\linewidth]{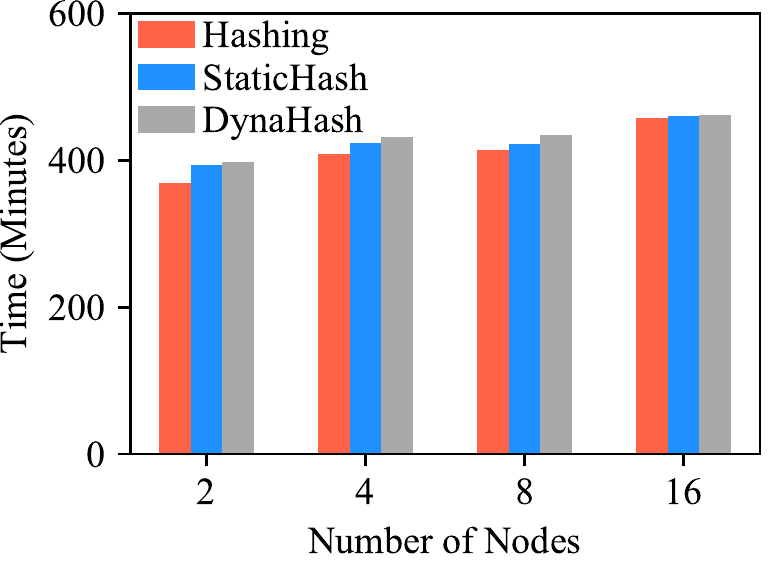}
	\caption{Ingestion Time}
	\label{fig:expr-ingestion}
\end{figure}

\begin{figure*}
	\centering
	\begin{subfigure}[b]{0.32\textwidth}
		\centering
		\includegraphics[width=\linewidth]{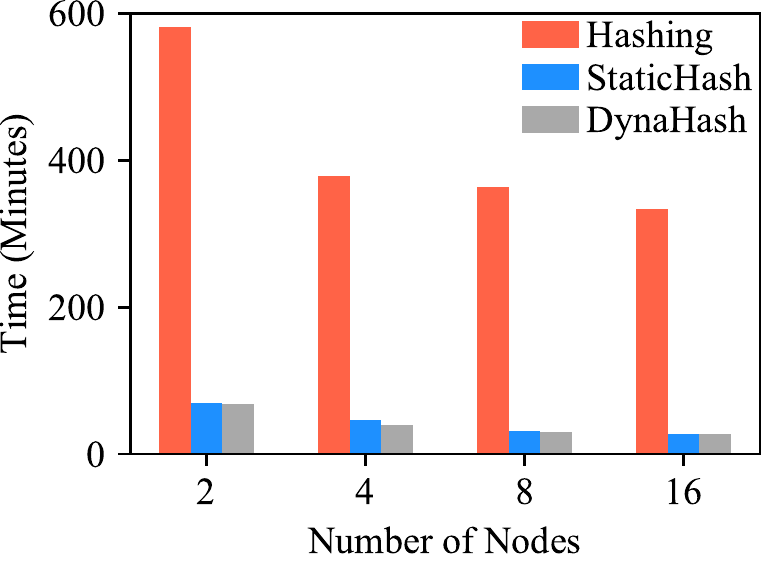}
		\caption{Removing Nodes}
		\label{fig:expr-rebalance-remove}
	\end{subfigure}
\hfil
	\begin{subfigure}[b]{0.32\textwidth}
		\centering
		\includegraphics[width=\linewidth]{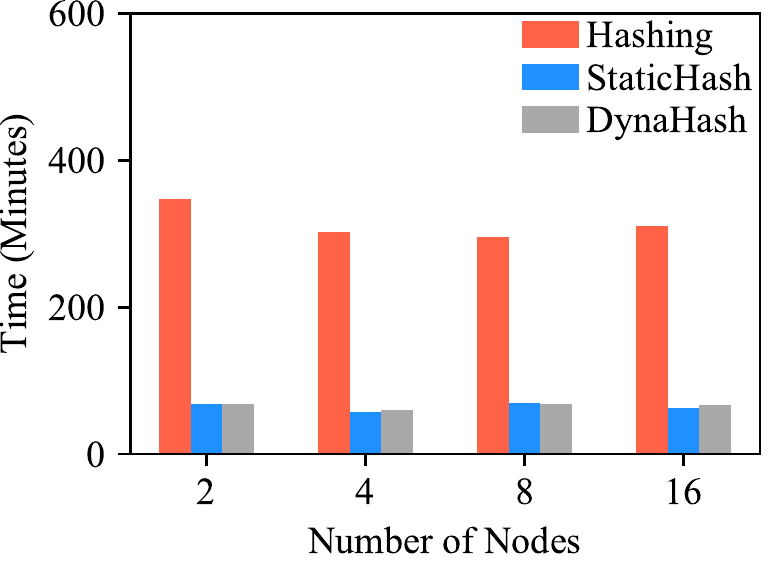}
		\caption{Adding Nodes}
		\label{fig:expr-rebalance-add}
	\end{subfigure}
\hfil
	\begin{subfigure}[b]{0.32\textwidth}
		\centering
		\includegraphics[width=\linewidth]{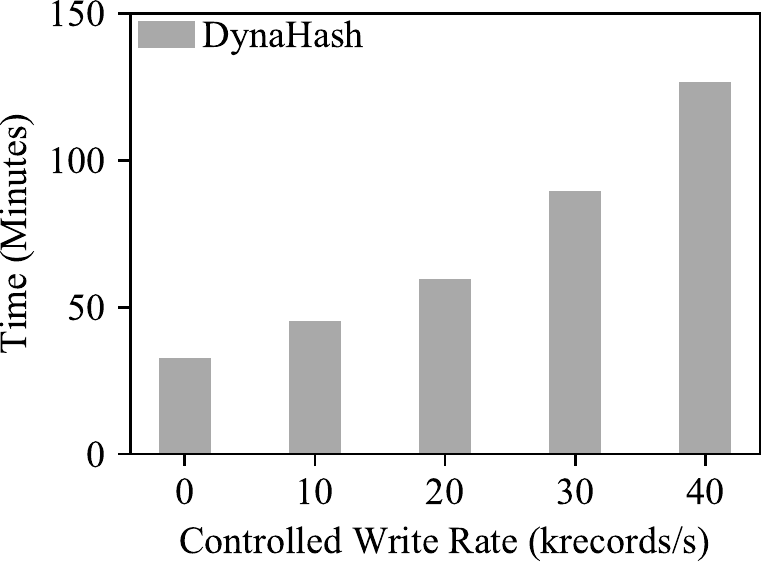}
		\caption{Concurrent Data Ingestion}
		\label{fig:expr-rebalance-write}
	\end{subfigure}
\caption{Rebalance Time}
\label{fig:expr-rebalance}
\end{figure*}

We first evaluated the ingestion performance of the different rebalancing approaches.
We used a TPC-H client that ran on a separate node to ingest all TPC-H data into the AsterixDB cluster.
The number of nodes of the AsterixDB cluster varied from 2 to 16.
The resulting ingestion time for each rebalancing approach under different cluster sizes is shown in \reffigure{fig:expr-ingestion}.
In general, \name incurs very a small overhead compared to AsterixDB's Hashing approach.
Moreover, comparing \name and \static, we see that
the different numbers of buckets per partition also have just a small impact on the ingestion performance.
When the cluster size increases, the ingestion time for all rebalancing
approaches slightly increases because of the write stall problem of LSM-trees:
In general, we found that data ingestion is relatively CPU-heavy in AsterixDB due to record parsing.
When a node has active merges, its ingestion rate will slow down due to the CPU contention caused by merges.
{This in turn will slow down the entire cluster because the overall performance of a shared-nothing system is bottlenecked by the slowest node, even though other nodes may not have ongoing merges.}
Thus, when the number of nodes increases, the write stall problem becomes more obvious, which increases the overall ingestion time.

\subsection{Rebalancing Performance}
Next we evaluated the rebalancing performance of the alternative rebalancing approaches, both for adding nodes and for removing nodes.
We further evaluated the impact of concurrent writes on the rebalance performance.

\begin{figure*}
	\centering
	\begin{subfigure}[b]{\textwidth}
		\centering
		\includegraphics[width=\size\linewidth]{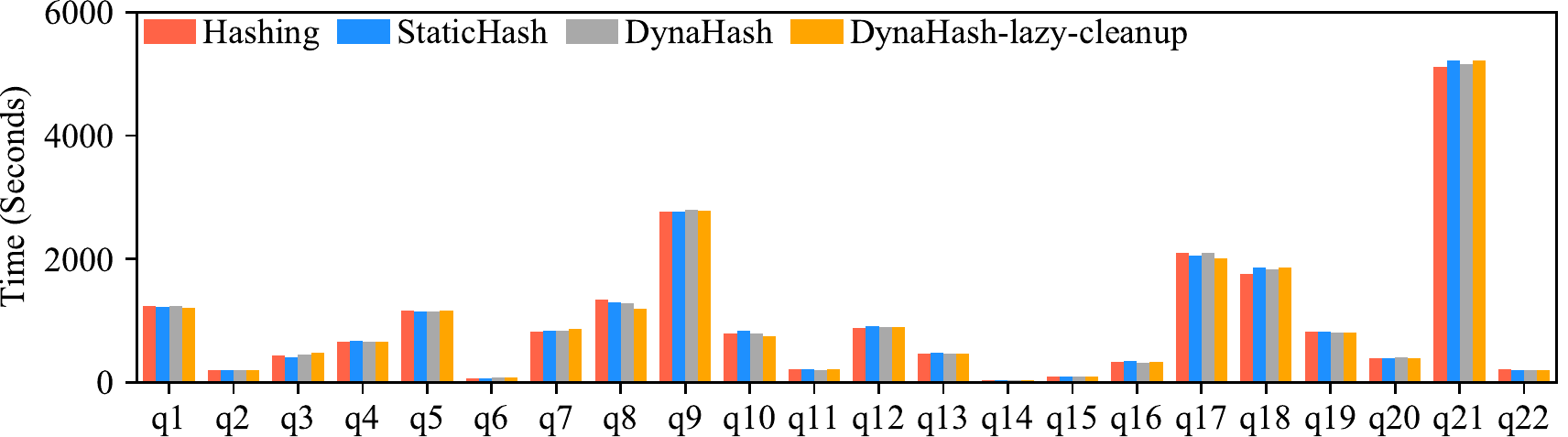}
		\caption{Query Performance on 4 Nodes}
		\label{fig:expr-query-4}
	\end{subfigure}
	\hfil
	\begin{subfigure}[b]{\textwidth}
		\centering
		\includegraphics[width=\size\linewidth]{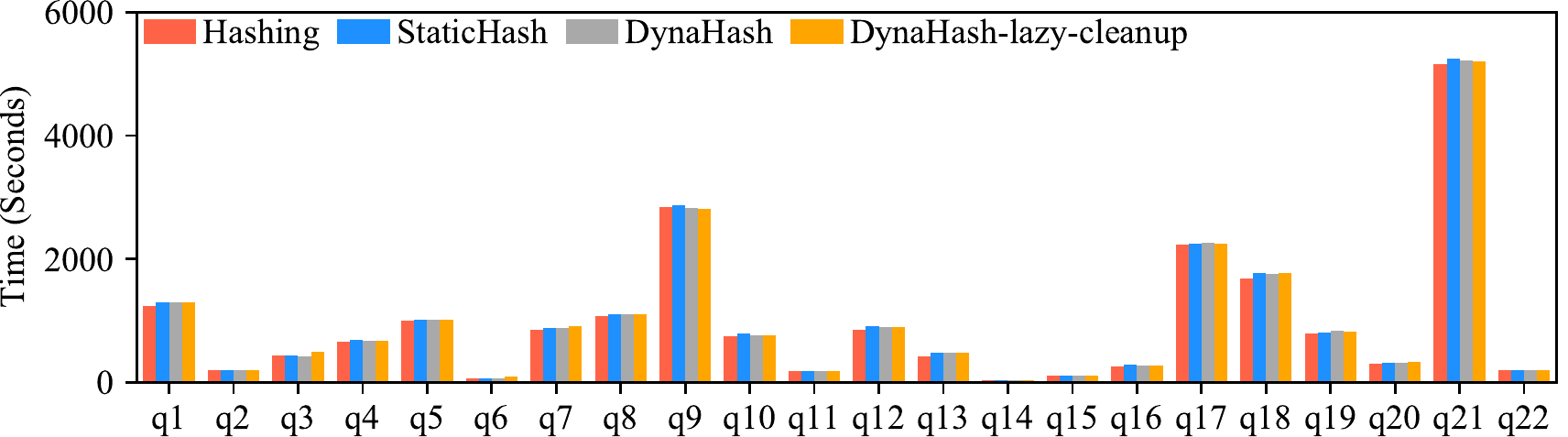}
		\caption{Query Performance on 16 Nodes}
		\label{fig:expr-query-16}
	\end{subfigure}
	\caption{Query Performance on Original Cluster}
	\label{fig:expr-query-original}
\end{figure*}

\textbf{Basic Rebalancing Performance.}
To understand the basic rebalancing performance of the different rebalancing approaches, we conducted the following experiments.
We first loaded the TPC-H datasets into an AsterixDB cluster with N nodes (N ranged from 2 to 16).
We then rebalanced all datasets to N-1 nodes to measure the time required to remove one node.
Finally, we rebalanced all datasets back to N nodes to measure the time required to add one node.

The rebalance times for removing and adding nodes are shown in \reffigure{fig:expr-rebalance}.
In general, both \static and \name substantially reduce the rebalancing time compared with Hashing for both removing and adding nodes.
Moreover, both of the bucketing approaches have similar rebalance times,
which shows that the number of buckets per partition has a small impact on the rebalancing performance.
Interestingly, we see that Hashing has different performance trends compared with \static and \name.
Hashing has better rebalancing performance for adding than for removing nodes
since its rebalancing work when adding is distributed across N nodes.
When a node is removed, however, the rebalancing work is largely distributed over N-1 nodes.
In contrast, for \static and \name,
removing a node is more efficient than adding one since the rebalancing work for node removal is distributed across the remaining N-1 nodes.
However, when a node is added, the new node becomes the bottleneck because it receives data from all N-1 existing nodes.

\textbf{Impact of Concurrent Writes.}
We further evaluated the impact of concurrent writes on the rebalancing performance of \name.
In this experiment, we rebalanced the datasets from 4 nodes to 3 nodes and inserted new records into the LineItem dataset while rebalancing was active.
We omitted StaticHash in this experiment. These two approaches had identical behavior because they had the same initial number of buckets
and bucket splitting was disabled during rebalancing.
The resulting rebalancing time under different write rates is shown in \reffigure{fig:expr-rebalance-write}.
As one can see, rebalance takes longer to finish when the write rate becomes larger.
This is expected because these concurrent writes compete for CPU and I/O resources with rebalance.
Thus, it is desirable to schedule rebalance operations during off-peak hours to minimize contention with user workloads.
However, as the result shows, even under high write rates, the rebalance operation can still be completed in a reasonable amount of time.

\begin{figure*}
	\centering
	\begin{subfigure}[t]{\textwidth}
		\centering
		\includegraphics[width=\size\linewidth]{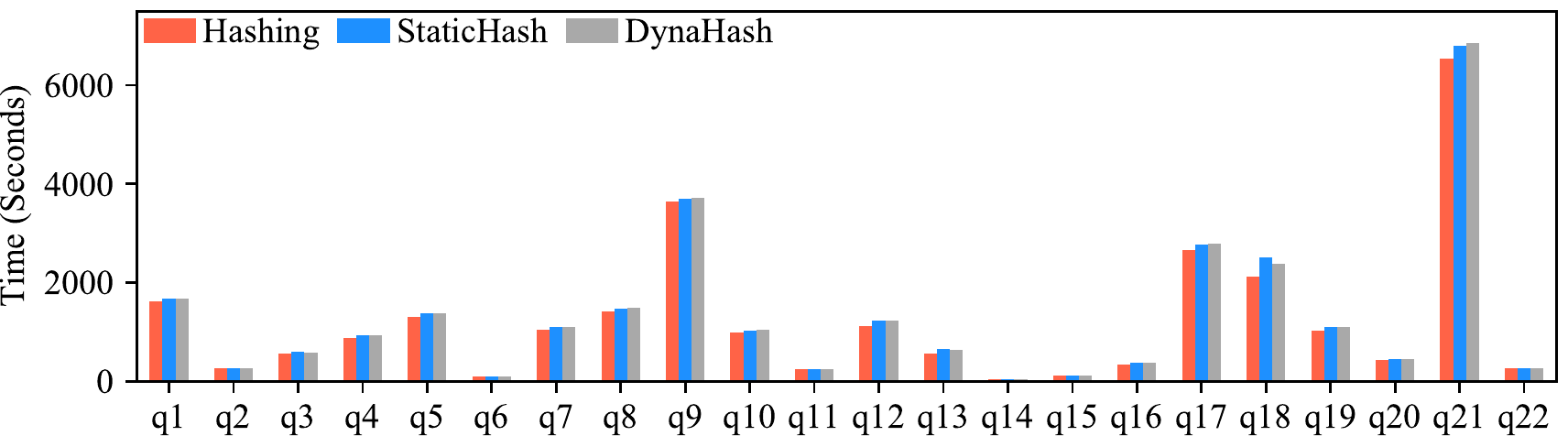}
		\caption{Query Performance on 3 Nodes}
		\label{fig:expr-query-3}
	\end{subfigure}
	\begin{subfigure}[t]{\textwidth}
		\centering
		\includegraphics[width=\size\linewidth]{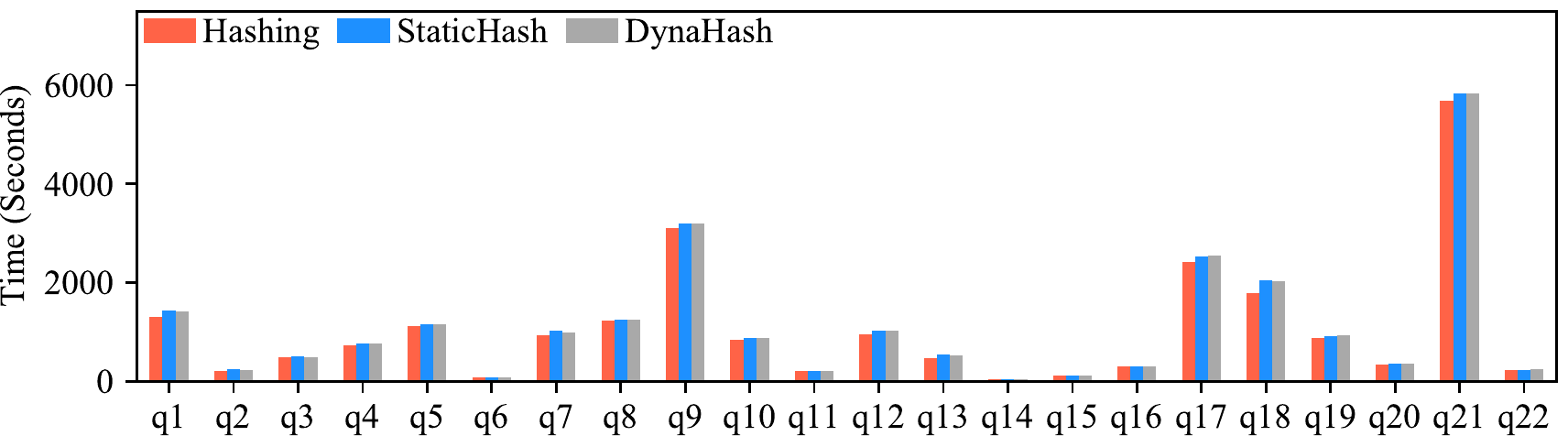}
		\caption{Query Performance on 15 Nodes}
		\label{fig:expr-query-15}
	\end{subfigure}
	\caption{Query Performance on Downsized Cluster}
	\label{fig:expr-query-resized}
\end{figure*}

\subsection{Query Performance}
Last but not least, we evaluated the OLAP query performance of the different rebalancing approaches, focusing on the following three questions:
First, what is the overhead of the proposed bucketed LSM-tree structure on query performance?
Second, what is the impact on query performance of the load balance resulting from the various rebalancing approaches?
Finally, what is the overhead due to lazy secondary index cleanup on query performance?

To answer these questions, we designed a series of experiments as follows.
First, we evaluated query performance on a cluster with 4 or 16 nodes without rebalancing, which helps to answer the first question.
We then rebalanced the datasets down to 3 or 15 nodes so that we can evaluate the load balance impact of the rebalancing approaches, which answers the second question.
Finally for \name, we rebalanced the datasets back to 4 or 16 nodes so that we can evaluate the performance impact of lazy secondary index cleanup (denoted as ``\name-lazy-cleanup").

The resulting query times on the original cluster size (4 or 16 nodes) are shown in \reffigure{fig:expr-query-original}.
(Recall that the data size is scaled in proportion to the cluster size.)
Note that on 16 nodes, \static and \name are expected to have similar behavior because they have the same number of buckets per partition.
In general, both \static and \name are seen to add a negligible overhead on most TPC-H queries when compared with Hashing.
This shows that bucketed LSM-trees have a negligible overhead for OLAP-style queries.
Moreover, all bucketing approaches achieve very good scale-up because the query times remain nearly constant when the number of nodes and dataset size increase.
Here one minor exception is q18, where \static and \name incur a small overhead compared with Hashing.
The reason is that q18 performs a groupby on the prefix of LineItem's primary keys, which requires the scanned records to be ordered on the primary keys.
In this case, the bucketed LSM-tree incurs some additional overhead because it has to merge-sort more disk components.
Moreover, \static also incurs a larger overhead on q18 under 4 nodes (\reffigure{fig:expr-query-4}) because it has 16 buckets per partition.
Finally, we see that lazy secondary index cleanup (\name-lazy-cleanup) also has a negligible overhead on TPC-H queries, which shows the effectiveness of this technique.
{With lazy secondary index cleanup, queries need to access some obsolete secondary index entries,
but the overhead is small compared to the overall query time.}

The resulting query times on the resized cluster (3 or 15 nodes) are shown in \reffigure{fig:expr-query-resized}.
Since the number of buckets cannot be evenly divided by the number of partitions,
both \static and \name result in some load imbalance where some partitions may have one more bucket than others.
Despite this load imbalance, both \static and \name only incur a very small overhead on most TPC-H queries.
This is because the load imbalance only impacts the data scan time, while most TPC-H queries are relatively computation heavy.
However, for scan-heavy queries, such as q17, q18, and q21, the overhead caused by a load imbalance is more noticeable.
For example, q17 and q18 each perform a full scan over the LineItem dataset to perform a groupby and aggregation, and q21 scans the LineItem dataset multiple times.
Moreover, as shown in \reffigure{fig:expr-query-3}, using more buckets per partition as in the \static approach slightly reduces
the overhead of the load imbalance for some queries, such as q21,
but doing so incurs some additional overhead for queries that require a sorted order coming from primary index scans, such as q18.

\subsection{Summary of Experimental Results}
In general, our experimental results are consistent with the discussion in \refsection{sec:rebalancing}.
Global rebalancing with hash partitioning achieves the best ingestion and query performance, but results in a very large rebalance cost.
In contrast, \name significantly reduces the rebalance time with only a small overhead on the ingestion and query performance.
The proposed bucketed LSM-tree structure only incurs a small overhead for queries that require the scanned
records to be ordered by primary keys.
Moreover, the load imbalance caused by \name mainly impacts the dataset scan performance.
Overall, \name imposes a negligible overhead on computation-intensive queries and just a small overhead on scan-heavy queries.
By comparing \static and \name, it can be seen that having more buckets per partition achieves a better load balance,
but it also leads to a larger overhead for queries that require the scanned records to be ordered on primary keys.
Moreover, \name requires less tuning since the number of buckets per partition is dynamically adjusted as the cluster and dataset size scales,
resulting in more stable performance.

\section{Conclusion}
\label{sec:conclusion}
In this paper, we have described the design and implementation of \name, an efficient data rebalancing approach that combines dynamic bucketing with extensible hashing in Apache AsterixDB.
We first introduced a bucketed LSM-tree design for efficiently storing multiple buckets.
We further described an efficient rebalancing implementation that exploits the LSM-tree's out-of-place update design to support concurrent reads and writes.
An experimental evaluation using the TPC-H benchmark has shown
that the proposed techniques significantly reduce the rebalance cost with negligible overheads for data ingestion and query processing.
{In the future, we plan to extend \name to incorporate replication to provide better availability and fault tolerance.}

\section*{Acknowledgment}
This work was supported by NSF awards CNS-1305430, IIS-1447720, IIS-1838248, and CNS-1925610 along with industrial support from Amazon, Google, and Microsoft and support from the Donald Bren Foundation (via a Bren Chair).

\balance

\bibliographystyle{IEEETranS}
\bibliography{rebalance.bib}

\end{document}